\pdfoutput=1

\documentclass[11pt]{article}

\usepackage{multirow}
\usepackage{graphicx}
\usepackage{booktabs}
\usepackage{tabularx}
\usepackage{caption}
\usepackage{amsmath}
\usepackage{svg}
\usepackage{afterpage}
\usepackage[review]{acl}

\usepackage{times}
\usepackage{latexsym}

\usepackage[T1]{fontenc}

\usepackage[utf8]{inputenc}

\usepackage{microtype}

\usepackage{inconsolata}

\usepackage[perpage]{footmisc}

\usepackage{graphicx}
\usepackage{tabularx}

\title{IRSC: A Zero-shot Evaluation Benchmark for Information Retrieval through Semantic Comprehension in Retrieval-Augmented Generation Scenarios}

\author{
 \textbf{Hai Lin\textsuperscript{1,2}\thanks{These authors contributed equally to this work.}},
 \textbf{Shaoxiong Zhan\textsuperscript{2}\footnotemark[1]},
 \textbf{Junyou Su\textsuperscript{3}\footnotemark[1]},
 \\
 \textbf{Haitao Zheng\textsuperscript{1,2}},
 \textbf{Hui Wang\textsuperscript{1}\thanks{Corresponding author.}}
\\
\\
 \textsuperscript{1}PengCheng Laboratory
 \\
 \textsuperscript{2}Shenzhen International Graduate School, Tsinghua University
 \\
 \textsuperscript{3}Southern University of Science and Technology
 \\
}

\begin{document}
\maketitle
\begin{abstract}
In Retrieval-Augmented Generation (RAG) tasks using Large Language Models (LLMs), the quality of retrieved information is critical to the final output. This paper introduces the IRSC benchmark for evaluating the performance of embedding models in multilingual RAG tasks. The benchmark encompasses five retrieval tasks: query retrieval, title retrieval, part-of-paragraph retrieval, keyword retrieval, and summary retrieval. Our research addresses the current lack of comprehensive testing and effective comparison methods for embedding models in RAG scenarios. We introduced new metrics: the Similarity of Semantic Comprehension Index (SSCI) and the Retrieval Capability Contest Index (RCCI), and evaluated models such as Snowflake-Arctic, BGE, GTE, and M3E. Our contributions include: 1) the IRSC benchmark, 2) the SSCI and RCCI metrics, and 3) insights into the cross-lingual limitations of embedding models. The IRSC benchmark aims to enhance the understanding and development of accurate retrieval systems in RAG tasks. All code and datasets are available at: https://github.com/Jasaxion/IRSC\_Benchmark
\end{abstract}

\section{Introduction}

The rapid advancements in large language models (LLMs) have demonstrated significant potential in natural language understanding and generation. However, these models still face challenges like factual hallucination, knowledge updating, and lack of domain-specific expertise \cite{chen2024benchmarking}. To address these issues, incorporating external knowledge through Retrieval-Augmented Generation (RAG) has emerged as a promising approach \cite{chen2024benchmarking, zhang2023miracl}.

RAG enhances LLMs by integrating retrieved information from external sources, which helps mitigate hallucinations and provide more accurate, up-to-date responses \cite{chen2024benchmarking}. Despite these advantages, existing benchmarks for evaluating RAG models are limited in scope and do not fully address the diverse needs of various retrieval tasks \cite{chen2024benchmarking,zhang2023miracl}. Most benchmarks focus primarily on tasks such as semantic textual similarity (STS), clustering, and re-ranking, but fail to provide RAG comprehensive evaluations across different retrieval scenarios.

The IRSC Benchmark introduced in this study aims to fill this gap by evaluating Embedding models across five distinct retrieval tasks: query-based retrieval, title-based retrieval, part-of-paragraph retrieval, keyword-based retrieval and summary-based retrieval. This benchmark is designed to reflect realistic application scenarios of RAG, considering different types of queries and languages (English, Chinese, and Mixed-Language datasets) \cite{zhang2023miracl}. We evaluate models such as Snowflake-Arctic-Embed-S \cite{merrick2024arcticembed}, BGE-M3 \cite{chen2024bge}, and \cite{MokaMassiveMixedEmbedding} across different tasks and languages, providing insights into their strengths and weaknesses in real-world RAG applications. Additionally, this benchmark includes innovative evaluation metrics to capture model performance differences across tasks and languages.

And due to the differences in vector dimensions and values across various models, directly computing cosine similarity between vectors \cite{steck2024cosine} is not feasible for comparing the semantic similarity between different models \cite{zhou2022problems}. To address this, we propose the Similarity of Semantic Comprehension Index (SSCI) in this paper. SSCI measures the similarity of semantic understanding between the model's output and the ground truth.

Our contributions are as follows:
1. We propose a comprehensive IRSC Benchmark to evaluate the performance of embedding models in RAG retrieval tasks and languages.
2. We introduce the SSCI and the Retrieval Capability Contest Index (RCCI) as innovative metrics to evaluate and compare models' semantic understanding and retrieval capabilities, respectively.
3. We conducted experiments on the retrieval effect of the model across languages and found the differences in the semantic understanding alignment of the model in different languages.

\section{Related Work}

The field of Retrieval-Augmented Generation (RAG) has gained significant attention, especially in addressing the limitations of Large Language models (LLMs) in providing accurate and contextually relevant information. This section reviews notable works in this domain and situates our contribution within the existing research.

\textbf{Benchmarking in RAG}

Chen et al. developed the Retrieval-Augmented Generation Benchmark (RGB) to evaluate LLMs on four abilities: noise robustness, negative rejection, information integration, and counterfactual robustness. Their findings highlight the need for nuanced evaluation metrics to improve RAG capabilities, as LLMs showed weaknesses in negative rejection, information integration, and handling false information\cite{chen2024benchmarking} . However, RGB primarily focuses on robustness aspects and does not provide comprehensive coverage of different retrieval tasks, which is crucial for real-world RAG applications.

\textbf{Multilingual Retrieval Datasets}

The MIRACL dataset, introduced by Zhang et al., supports multilingual information retrieval with 700,000 human-annotated query-passage pairs across 18 languages. It aims to advance retrieval models that handle linguistic diversity and resource variability \cite{zhang2023miracl}. While MIRACL provides valuable multilingual data, it is mainly focused on query-passage retrieval and does not address other important retrieval tasks like keyword or title retrieval.

\textbf{Evaluations of RAG Systems}

Ogundepo et al. provided a comprehensive survey of current evaluation methods for RAG systems, emphasizing the importance of various retrieval tasks and metrics like nDCG, MRR, and MAP \cite{yu2024evaluation}. Their work discusses challenges and future directions for robust RAG benchmarks. Despite their comprehensive survey, there is a lack of practical benchmarks that integrate these varied metrics across different retrieval tasks.

\textbf{Benchmark for Evaluation of Information Retrieval Models (BEIR)}

Thakur et al. introduced an evaluation benchmark for retrieval models called BEIR, which includes a diverse set of information retrieval tasks across different domains and data types\cite{thakur2021beir}. BEIR offers a collection of heterogeneous tasks and provides a unified and convenient framework for evaluating retrieval models based on natural language processing. However, BEIR only focuses on retrieval tasks between queries and paragraphs, and does not address the more complex retrieval tasks involving large-scale RAG (Retrieval-Augmented Generation) scenarios.

\textbf{Massive Text Embedding Benchmark (MTEB)}

Muennighoff et al. introduced MTEB, a benchmark evaluating text embedding models across tasks such as bitext mining, classification, clustering, reranking, retrieval, and semantic textual similarity\cite{muennighoff2023mteb} . Their findings highlight the need for specialized models tailored to specific retrieval scenarios. However, MTEB does not focus specifically on the integration of retrieved information for generation tasks, which is a critical component of RAG systems.

\textbf{Multilingual Question Answering}

The MKQA dataset, presented by Longpre et al., evaluates multilingual open-domain question answering systems with parallel questions in multiple languages\cite{longpre2021mkqa}. It facilitates comparative analysis of retrieval and QA performance across different linguistic contexts. While useful for question answering, MKQA does not encompass the broader spectrum of retrieval tasks that are essential for RAG evaluations.

\textbf{Current Work on RAG Model Evaluation}

Our work extends these studies by proposing a novel Benchmark that evaluates retrieval performance across five tasks: query, keyword, title, summary, and part of paragraph retrieval. Unlike previous benchmarks, our dataset includes multilingual and cross-lingual components, addressing the need for robust evaluation in diverse linguistic environments. Our benchmark aims to fill gaps in existing methods by providing a comprehensive assessment of model performance in RAG tasks, focusing on cross-lingual retrieval and integrating various retrieval tasks into a unified framework.

\section{The IRSC Benchmark}

\subsection{Desiderata}

The IRSC benchmark is designed to evaluate the effectiveness of embedding models specifically within the context of Retrieval-Augmented Generation (RAG) tasks. Unlike traditional benchmarks that focus broadly on sentence or paragraph length, IRSC hones in on the unique needs of RAG applications, which require supplementing knowledge to queries. This benchmark emphasizes five key data types to cover most RAG tasks:

\begin{enumerate}
    \item \textbf{Focus on RAG-Specific Retrieval Tasks:} Unlike traditional benchmarks, IRSC focuses on expanding a query or brief information into a detailed response.
    \item \textbf{Emphasis on Cross-lingual Capabilities:} IRSC evaluates models in multiple languages, particularly English and Chinese, to handle Mixed-Language queries and adapt to cross-lingual environments.
    \item \textbf{Comprehensive Evaluation Metrics:} Standard retrieval metrics (nDCG@10, MRR@10, MAP@10, precision@3, and recall@10) are used alongside new metrics like SSCI and RCCI for deeper insights into semantic comprehension and retrieval capabilities.
    \item \textbf{Real-World Applicability:} IRSC focuses on real-world RAG tasks like retrieving detailed knowledge based on a query or summary, ensuring practical relevance and cross-lingual applicability.
\end{enumerate}

Through these considerations, IRSC aims to set a new standard for evaluating embedding models in the context of RAG tasks, providing a more nuanced and applicable assessment framework.

\begin{figure}[h]
    \centering
    \resizebox{0.95\linewidth}{!}{\includegraphics{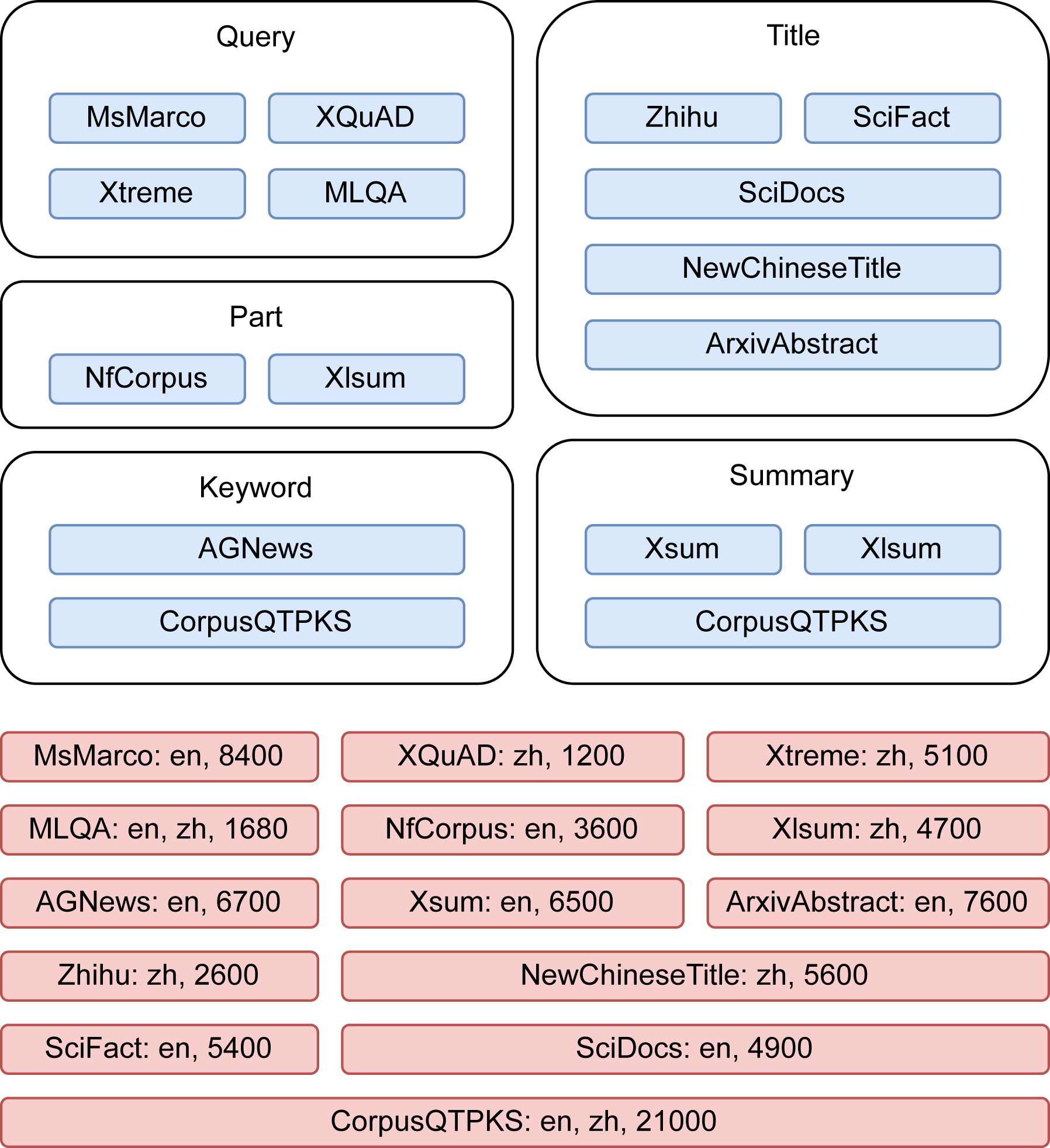}}
    \caption{The IRSC Benchmark is structured around five primary task types, each designed to evaluate different aspects of a model's retrieval capabilities. The red labels indicate the languages and quantities of each dataset.}
    \label{fig:Overview}
\end{figure}

\subsection{Tasks and Evaluation}
Figure~\ref{fig:Overview} provides an overview of tasks and datasets available in IRSC. The benchmark consists of the following five task types:

\begin{enumerate}

    \item \textbf{Query -> Paragraph:} Evaluates the model's ability to retrieve relevant paragraphs based on a given query. Datasets: MsMARCO\cite{bajaj2018ms}, XQuAD\cite{Artetxe_2020}, Xtreme\cite{hu2020xtreme}, and MLQA\cite{lewis2020mlqa}.
    
    \item \textbf{Title -> Paragraph:} Tests the model's capability to find relevant paragraphs given a title. Datasets: zhihu\footnote{\url{https://huggingface.co/datasets/suolyer/zhihu}}, New-Title-Chinese\footnote{\url{https://huggingface.co/datasets/madao33/new-title-chinese}}, Arxiv-Abstract\cite{clement2019use}, SciDocs\cite{muennighoff2023mteb}, and SciFact\cite{muennighoff2023mteb}.
    
    \item \textbf{Part of Paragraph -> Paragraph:} Evaluates sensitivity to text fragments, testing if the model can retrieve the full paragraph from a fragment. Datasets: nfcorpus\cite{muennighoff2023mteb} (English) and xlsum\cite{joulin2017bag} (Chinese).
    
    \item \textbf{Keyword -> Paragraph:} Measures the model's ability to retrieve paragraphs based on keywords. Datasets: AG News\cite{zhang2015character} and CorpusQTPKS.
    
    \item \textbf{Summary -> Paragraph:} Evaluates performance in retrieving relevant paragraphs based on a summary. Datasets: XSum\cite{narayan2018don} (English), xlsum\cite{joulin2017bag} (Chinese), and CorpusQTPKS.
\end{enumerate}

Each task uses 5000 query-content pairs as evaluation data. The remaining samples form a unified database for retrieval across all tasks. During scoring, queries are used to search the unified database, and retrieval performance is evaluated based on the precision of retrieved indices against the ground truth. This standardized approach ensures robust and fair evaluation of the model's retrieval capabilities across various tasks.

\subsection{Evaluation Metrics}
The primary evaluation metrics for IRSC include nDCG@10, MRR@10, MAP@10, precision@3, and recall@10. These metrics are used to evaluate model performance in information retrieval tasks:

\textbf{Recall@10:} Evaluates the fraction of relevant documents retrieved among the top 10 documents.

\textbf{MRR@10 (Mean Reciprocal Rank):} Evaluates the rank position of the first relevant document.

\textbf{nDCG@10 (Normalized Discounted Cumulative Gain):} Measures the ranking quality of the retrieved documents, taking into account the position of relevant documents in the ranking.

In addition to these standard metrics, we introduce new metrics to assess different aspects of model performance:

    

\subsubsection*{Similarity of Semantic Comprehension Index (SSCI)}

The Similarity of Semantic Comprehension Index, averaged over multiple queries. It measures the difference in semantic understanding between the two models' outputs across all queries. A higher value indicates a greater disparity in the models' understanding of the given questions.
    
We define the average SSCI (\(\overline{\text{SSCI}}\)) as:
    \[
    \overline{\text{SSCI}} = \frac{1}{Q} \sum_{q=1}^{Q} \frac{|m1_q - m2_q|}{n}
    \]

\subsubsection*{Retrieval Capability Contest Index (RCCI)}

The Retrieval Capability Contest Index, averaged over multiple queries. It evaluates the differences in retrieval capabilities between the two models across all queries. A positive average score indicates that model 1 performs better overall, while a negative average score indicates that model 2 performs better overall. The magnitude of the score indicates the extent of the difference in performance.
    
We define the average RCCI (\(\overline{\text{RCCI}}\)) as:
    \[
    \overline{\text{RCCI}} = \frac{1}{Q} \sum_{q=1}^{Q} \frac{m1_q - m2_q}{n}
    \]

\subsubsection*{Parameters}

\begin{itemize}
    \item \( n \): The length of each query result vector minus one, representing the maximum index of the retrieval results.
\end{itemize}

\begin{itemize}
    \item \( R1, R2 \): These are matrices representing the results retrieved by the two different models over \( Q \) queries. Each matrix has dimensions \( Q \times (n+1) \). Each element is a binary value (0 or 1), where 1 indicates the position of the correct answer for each query.
    
    For query \( q \), the vectors \( R1_q \) and \( R2_q \) are defined as follows:
    \[
    R1_q = [r_{11q}, r_{12q}, r_{13q}, \ldots, r_{1nq}, r_{1(n+1)q}]
    \]
    \[
    R2_q = [r_{21q}, r_{22q}, r_{23q}, \ldots, r_{2nq}, r_{2(n+1)q}]
    \]
\end{itemize}

\begin{itemize}
    \item \( m1_q, m2_q \): These represent the positions of the correct answer in \( R1_q \) and \( R2_q \) respectively for query \( q \). If there is no 1 in the vector, it is assigned a value of -1.
    
    For query \( q \):
    \[
    m1_q = 
    \begin{cases} 
    n - \text{index}(R1_q, 1) & \text{if } 1 \in R1_q \\
    -1 & \text{otherwise}
    \end{cases}
    \]
    \[
    m2_q = 
    \begin{cases} 
    n - \text{index}(R2_q, 1) & \text{if } 1 \in R2_q \\
    -1 & \text{otherwise}
    \end{cases}
    \]
    
    where \( \text{index}(R_q, 1) \) denotes the index position of the element equal to 1 in the vector \( R_q \).
\end{itemize}

By using these metrics, IRSC aims to provide a comprehensive evaluation framework for assessing the performance of embedding models across diverse retrieval tasks.

\subsection{Model Descriptions}

We evaluated 13 models using the IRSC Benchmark. Notably, \texttt{MiniLM-L6-v2} models do not support Chinese. 

\begin{itemize}
    \item \textbf{S-Arctic Series}\cite{merrick2024arcticembed}: Includes \texttt{S-Arctic-S}, \texttt{S-Arctic-M}, and \texttt{S-Arctic-L}, designed for semantic embeddings in text retrieval tasks.
    
    \item \textbf{BGE Series}\cite{chen2024bge}: Includes \texttt{BGE-M3} (multilingual) and \texttt{BGE-Large} (optimized for Chinese).
    
    \item \textbf{GTE Series}\cite{li2023towards}: Comprises \texttt{GTE-Small}, \texttt{GTE-Base}, and \texttt{GTE-Large}, focusing on general text embeddings.
    
    \item \textbf{M3E Series}\cite{MokaMassiveMixedEmbedding}: Includes \texttt{M3E-Small}, \texttt{M3E-Base}, and \texttt{M3E-Large}, designed for efficient multilingual text embeddings.
    
    \item \textbf{MiniLM Series}: MiniLM-L12\cite{reimers-2019-sentence-bert}: A multilingual version of the MiniLM series, tailored for paraphrase identification and multilingual retrieval. MiniLM-L6\footnote{\url{https://huggingface.co/sentence-transformers/MiniLM-L6-v2}}: A compact, efficient model focused on English for various NLP tasks.
\end{itemize}

\begin{table*}
  \centering
  \resizebox{\textwidth}{!}{
    \begin{tabular}{lccccccccccccccc}
      \toprule
      \multirow{2}{*}{Model} & \multicolumn{3}{c}{Query} & \multicolumn{3}{c}{Title} & \multicolumn{3}{c}{Part} & \multicolumn{3}{c}{Keywords} & \multicolumn{3}{c}{Summary} \\
      \cmidrule(lr){2-4} \cmidrule(lr){5-7} \cmidrule(lr){8-10} \cmidrule(lr){11-13} \cmidrule(lr){14-16}
      & r@10 & m@10 & n@10 & r@10 & m@10 & n@10 & r@10 & m@10 & n@10 & r@10 & m@10 & n@10 & r@10 & m@10 & n@10 \\
      \midrule
        S-Arctic-S & 0.3067 & 0.2714 & 0.2815 & 0.3566 & 0.3125 & 0.3232 & 0.4588 & 0.4369 & 0.4423 & 0.6302 & 0.5759 & 0.5892 & 0.5334 & 0.5231 & 0.5256 \\ 
        S-Arctic-M & 0.1379 & 0.1125 & 0.1196 & 0.0198 & 0.0145 & 0.0158 & 0.2746 & 0.2514 & 0.2570 & 0.2856 & 0.2339 & 0.2464 & 0.4554 & 0.4151 & 0.4247 \\
        S-Arctic-L & 0.2238 & 0.1909 & 0.2002 & 0.0248 & 0.0186 & 0.0200 & 0.3126 & 0.2903 & 0.2957 & 0.3804 & 0.3342 & 0.3455 & 0.4394 & 0.4104 & 0.4175 \\
        BGE-M3 & \textbf{0.6972} & \textbf{0.6321} & \textbf{0.6495} & \textbf{0.8640} & \textbf{0.8149} & \textbf{0.8270} & \textbf{0.7964} & \textbf{0.7625} & \textbf{0.7708} & \textbf{0.8668} & \textbf{0.8205} & \textbf{0.8320} & \textbf{0.9812} & \textbf{0.9709} & \textbf{0.9735} \\
        GTE-Small & 0.5099 & 0.4643 & 0.4771 & 0.7360 & 0.7005 & 0.7093 & 0.6916 & 0.6527 & 0.6622 & 0.7876 & 0.7258 & 0.7409 & 0.8284 & 0.7890 & 0.7986 \\ 
        GTE-Base & 0.5163 & 0.4684 & 0.4817 & 0.7366 & 0.7027 & 0.7110 & 0.6980 & 0.6617 & 0.6706 & 0.7940 & 0.7333 & 0.7482 & 0.8282 & 0.7893 & 0.7987 \\ 
        GTE-Large & 0.5205 & 0.4746 & 0.4874 & 0.7372 & 0.7022 & 0.7107 & 0.6984 & 0.6571 & 0.6672 & 0.7936 & 0.7338 & 0.7485 & 0.8180 & 0.7780 & 0.7877 \\
        M3E-Small & 0.2292 & 0.1874 & 0.1981 & 0.2850 & 0.2267 & 0.2407 & 0.4942 & 0.4522 & 0.4624 & 0.3374 & 0.2874 & 0.2993 & 0.8052 & 0.7572 & 0.7689 \\ 
        M3E-Base & 0.5912 & 0.5239 & 0.5415 & 0.7840 & 0.7167 & 0.7332 & 0.7562 & 0.7146 & 0.7249 & 0.8368 & 0.7777 & 0.7923 & 0.9644 & 0.9441 & 0.9491 \\
        M3E-Large & 0.3415 & 0.2798 & 0.2957 & 0.5052 & 0.4113 & 0.4340 & 0.5788 & 0.5302 & 0.5421 & 0.5606 & 0.4701 & 0.4919 & 0.8964 & 0.8527 & 0.8634 \\
        MiniLM-L6 & 0.4589 & 0.4180 & 0.4297 & 0.6168 & 0.5842 & 0.5923 & 0.6066 & 0.5720 & 0.5805 & 0.7042 & 0.6326 & 0.6500 & 0.5484 & 0.5190 & 0.5260 \\
        MiniLM-L12 & 0.4934 & 0.4161 & 0.4363 & 0.6174 & 0.5294 & 0.5508 & 0.5708 & 0.5259 & 0.5368 & 0.5784 & 0.4846 & 0.5073 & 0.8728 & 0.8245 & 0.8365 \\
      \hline
    \end{tabular}
  }
  \caption{\label{all-language-result} IRSC Benchmark Results of S-Arctic Series, BGE Series, GTE Series, M3E Series, and MiniLM Series in All Languages for All Tasks. \textbf{Metrics:} r@10 - Recall at 10, m@10 - MRR(Mean Reciprocal Rank) at 10, n@10 - nDCG(Normalized Discounted Cumulative Gain) at 10}
\end{table*}

\begin{figure*}[h]
    \centering
    \resizebox{0.95\linewidth}{!}{\includegraphics{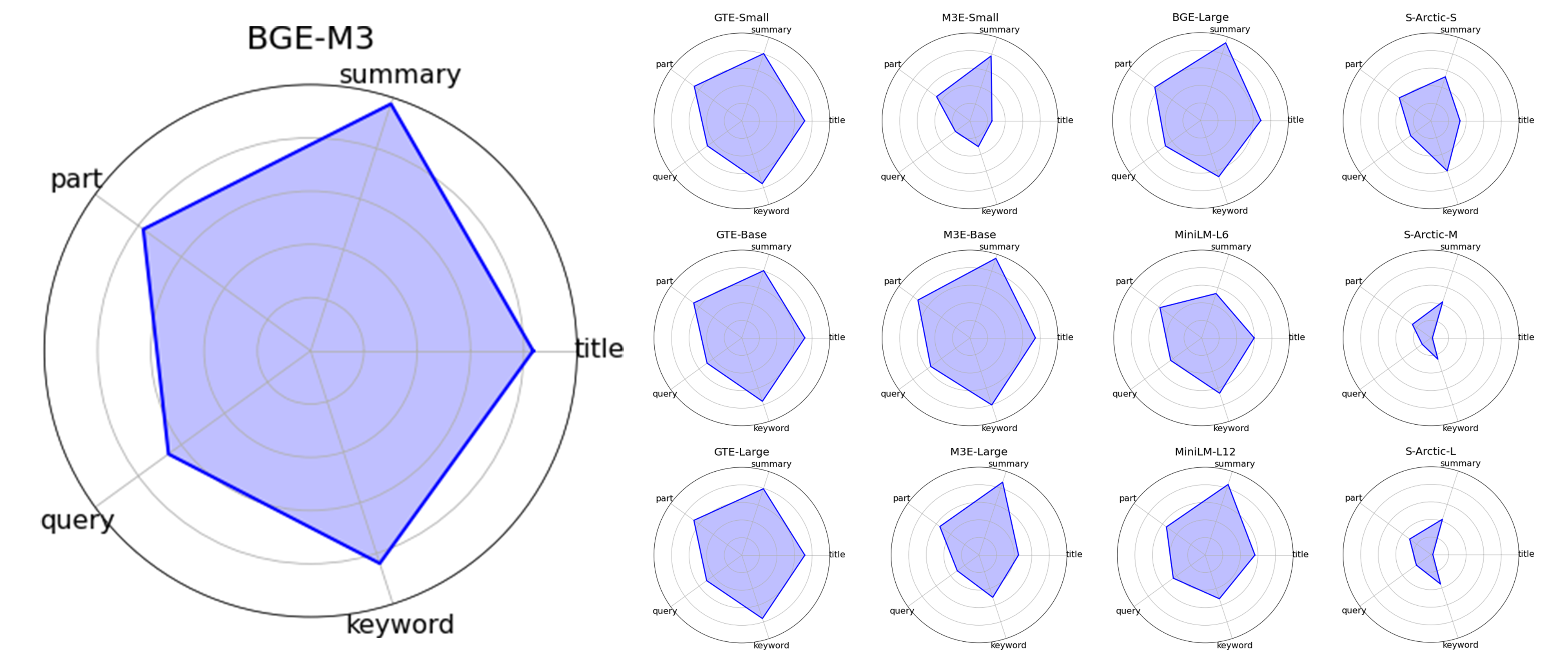}}
    \caption{Comparative Performance Radar Charts of S-Arctic Series, BGE Series, GTE Series, M3E Series, and MiniLM Series Models Across IRSC Benchmark's Query, Title, Part, Keyword and Summary Tasks in Mixed-Language. \textbf{Metrics:} Average of Recall@10, MRR@10 and nDCG@10}
    \label{fig:radar}
\end{figure*}

\section{Results}

\subsection{Experimental Setup}

The experiments are conducted across three different language requirements: English, Chinese, and Mixed-Language (English + Chinese). For each language requirement, corresponding benchmark datasets are utilized to perform IRSC scoring experiments.

\textbf{English}: The evaluation involves English-specific benchmark datasets to test the retrieval performance of each model.

\textbf{Chinese}: The evaluation uses Chinese-specific benchmark datasets, ensuring the models' capabilities are tested in the Chinese language context.

\textbf{Mixed-Language (English + Chinese)}: This mixed evaluation assesses the models' performance across both English and Chinese datasets, providing a comprehensive understanding of their cross-lingual retrieval capabilities.

By employing this diversified language setup, we aim to provide a thorough and robust evaluation of each model's performance in retrieving relevant paragraphs based on various query types within the IRSC benchmark.

\subsection{Experimental Results and Analysis}

\subsubsection{Benchmark Analysis}

Based on the results in Table~\ref{all-language-result}, we observe that the BGE-M3 model consistently outperforms other models across all metrics and categories, indicating its robustness and effectiveness in both Chinese and English retrieval tasks. Specifically, BGE-M3 achieves the highest recall at 10 (r@10), mean average precision at 10 (m@10), and normalized discounted cumulative gain at 10 (n@10) in the Keywords, Title, Query, Part, and Summary categories. For instance, in the Keywords category, BGE-M3 has an impressive r@10 of 0.8668, m@10 of 0.8205, and n@10 of 0.8320.

Conversely, the S-Arctic series (S-Arctic-S, S-Arctic-M, S-Arctic-L) shows relatively lower performance compared to other models. Notably, S-Arctic-S performs better than S-Arctic-M and S-Arctic-L across most categories, but it still lags significantly behind models like BGE-M3, GTE-Small, and M3E-Base. For example, in the Summary category, S-Arctic-S achieves an r@10 of 0.5334, whereas BGE-M3 achieves an r@10 of 0.9812.

Models like GTE-Base and M3E-Base also demonstrate strong performance, particularly in the Keywords and Summary categories. GTE-Base achieves an r@10 of 0.7940 in the Keywords category and M3E-Base achieves an r@10 of 0.9644 in the Summary category, showing their potential effectiveness in specific retrieval contexts.

We observe an interesting performance pattern between the S-Arctic-S and M3E-Small models. Notably, S-Arctic-S performs significantly better than M3E-Small in the Keywords category. S-Arctic-S achieves an r@10 of 0.6302, while M3E-Small scores 0.3374. However, in Summary categories, M3E-Small significantly outperforms S-Arctic-S. M3E-Small achieves an r@10 of 0.8052 compared to S-Arctic-S's 0.5334. This pattern indicates that while S-Arctic-S excels in Keywords retrieval, it falls behind in other tasks such as Summary, where M3E-Small demonstrates superior performance.

The diverse performance across different models highlights the importance of selecting the appropriate model based on the specific retrieval task and the language requirements. Table~\ref{all-language-result} presents the results for Mixed-Language task. The results for the Chinese and English tasks will publish in our Github repository.

\subsubsection{Radar Chart Analysis}

To more intuitively showcase the performance of different models, we created radar charts\ref{fig:radar} where the values for each capability are derived from the average of \(r@10\), \(m@10\), and \(n@10\). These charts provide a clearer view of the comprehensive performance of each model across various tasks.

From the charts, it is evident that BGE-M3 performs exceptionally well in all tasks (Query, Title, Part, Keywords, and Summary), demonstrating its comprehensive advantages across these five areas. The radar chart for BGE-M3 shows a balanced and extensive coverage, with particularly outstanding performance in the Summary tasks.

In contrast, the GTE series and M3E series models also show good performance. However, the S-Arctic series underperforms compared to the aforementioned models in all tasks, especially S-Arctic-M, which shows the lowest comprehensive performance across all tasks, indicating its lesser effectiveness in these tasks.

The radar charts clearly illustrate the comprehensive capabilities of each model across different tasks, with BGE-M3 standing out as the most optimal model in terms of performance.

\subsubsection{Cross Language Analysis}

\begin{table}[h]
  \centering
  \begin{tabular}{lcc}
    \hline
    \textbf{Model}  & \textbf{C2C} | \textbf{C2E} & \textbf{E2E} | \textbf{E2C} \\
    \hline
        S-Arctic-S & 0.0782 | 0.0068 & 0.5848 | 0.0462 \\
        S-Arctic-M & 0.1272 | 0.0014 & 0.1441 | 0.0208 \\
        S-Arctic-L & 0.0882 | 0.0008 & 0.2008 | 0.0334 \\
        BGE-M3 & 0.8630 | 0.6260 & 0.8427 | 0.5964 \\
        GTE-Small & 0.4088 | 0.0569 & 0.8499 | 0.0620 \\
        GTE-Base & 0.4048 | 0.0581 & 0.8613 | 0.0866 \\
        GTE-Large & 0.4036 | 0.0651 & 0.8693 | 0.0888 \\
        M3E-Small & 0.7486 | 0.0660 & 0.1327 | 0.0190 \\
        M3E-Base & 0.8026 | 0.3323 & 0.7423 | 0.1578 \\
        M3E-Large & 0.7648 | 0.2420 & 0.3659 | 0.0688 \\
        MiniLM-L6 & 0.1048 | 0.0209 & 0.7942 | 0.0150 \\
        MiniLM-L12 & 0.5586 | 0.3841 & 0.5872 | 0.4558 \\ \hline
  \end{tabular}
  \caption{\label{tab:cross-language-result} IRSC Benchmark Results of the S-Arctic Series, BGE Series, GTE Series, M3E Series, and MiniLM Series in Cross Languages. \textbf{Metrics:} recall@10}

\end{table}

In  Table~\ref{tab:cross-language-result} , we also conducted experiments on the cross-lingual retrieval capabilities of different models using five IRSC tasks, with 1,000 randomly selected queries for each task. We obtained 5,000 data entries in both English and Chinese languages. Queries originally in English were translated into the target language (Chinese) and then searched within an entirely English database to obtain the Chinese to English (C2E) results in Table~\ref{tab:cross-language-result}. The scores are the averages of r@10, m@10, and n@10.

From the results, several key observations can be made:
\begin{enumerate}
     \item \textbf{Performance Decline in Cross-Lingual Retrieval}: Most models exhibit a decline in performance metrics when transitioning from monolingual (C2C or E2E) to cross-lingual (C2E or E2C) retrieval. This indicates a general challenge in maintaining semantic alignment across different languages.
     \item \textbf{Superior Performance of BGE-M3}: The BGE-M3 model consistently demonstrates superior performance in both monolingual and cross-lingual retrieval tasks. Notably, its performance degradation from monolingual to cross-lingual retrieval is minimal. For instance, in C2C, it scores 0.8630, while in C2E, it scores 0.6260. Similarly, in E2E, it scores 0.8427, compared to 0.5964 in E2C.
     \item \textbf{Significant Decline in M3E Series}: The M3E series models show a significant decrease in performance when moving to cross-lingual tasks. The most notable drop is observed in the M3E-Base model, which falls from 0.8026 in C2C to 0.3323 in C2E. This highlights a substantial challenge in the model's ability to align queries semantically across languages.
     \item \textbf{Drastic Decline in GTE Series}: The GTE series models exhibit the most drastic decline in performance, especially in the E2C task. Scores around 0.85 in E2E drop below 0.1 in E2C, indicating a significant deficiency in the models' ability to handle cross-lingual semantic alignment from English to Chinese.
     \item \textbf{Mixed Performance in S-Arctic and MiniLM Series}: The S-Arctic series models display varying levels of performance, with S-Arctic-S and S-Arctic-M performing poorly in C2E tasks. The MiniLM series also shows mixed results, with MiniLM-L12 performing relatively well compared to MiniLM-L6 in cross-lingual tasks.

\end{enumerate}

These findings underscore the need for further improvement in training vector models for cross-lingual query semantic alignment. Enhancing the models' ability to maintain semantic coherence across languages could lead to more effective and accurate cross-lingual retrieval systems. Future research should focus on developing techniques to bridge the semantic gap between languages, ensuring that models can perform consistently well in both monolingual and cross-lingual contexts.

\subsection{SSCI \& RCCI Analysis}

\subsubsection{SSCI}

\begin{figure}[h]
    \centering
    \resizebox{1\linewidth}{!}{\includegraphics[]{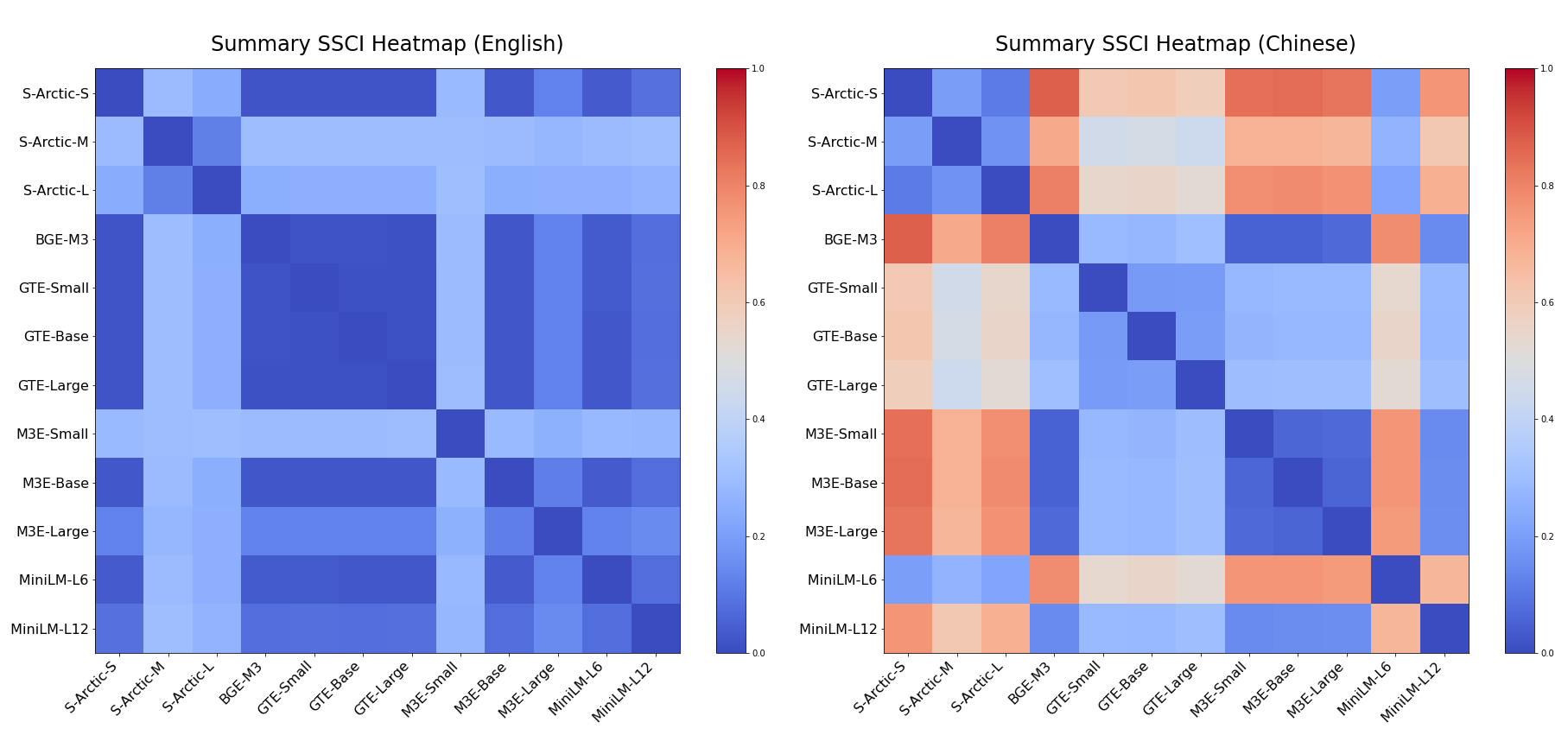}}
    \caption{Comparative SSCI Heatmaps of the S-Arctic Series, BGE Series, GTE Series, M3E Series, and MiniLM Series in the IRSC Benchmark's Summary Subtask Across Chinese and English. Smaller values indicate more consistent model performance.}
    \label{figure:ssci}
\end{figure}

In Figure~\ref{figure:ssci}, we present detailed SSCI results for the Summary task across different languages and models. We observe that in the English language, most models display blue regions, indicating high consistency in semantic understanding among these models. In contrast, for the Chinese language, the SSCI values exhibit more red regions, suggesting lower consistency and greater divergence in semantic understanding among the models. 

Furthermore, by examining the color distribution in Figure~\ref{figure:ssci}, we find that models within the same series generally exhibit better semantic understanding consistency, whereas models from different series are more likely to show divergence in understanding. From this analysis, we can draw several conclusions: there is a significant difference in semantic understanding consistency across languages, with models showing higher consistency in English compared to Chinese; models within the same series tend to have higher semantic understanding consistency, while different series of models are more prone to divergences in understanding.

\begin{figure}[h]
    \centering
    \resizebox{1\linewidth}{!}{\includegraphics[]{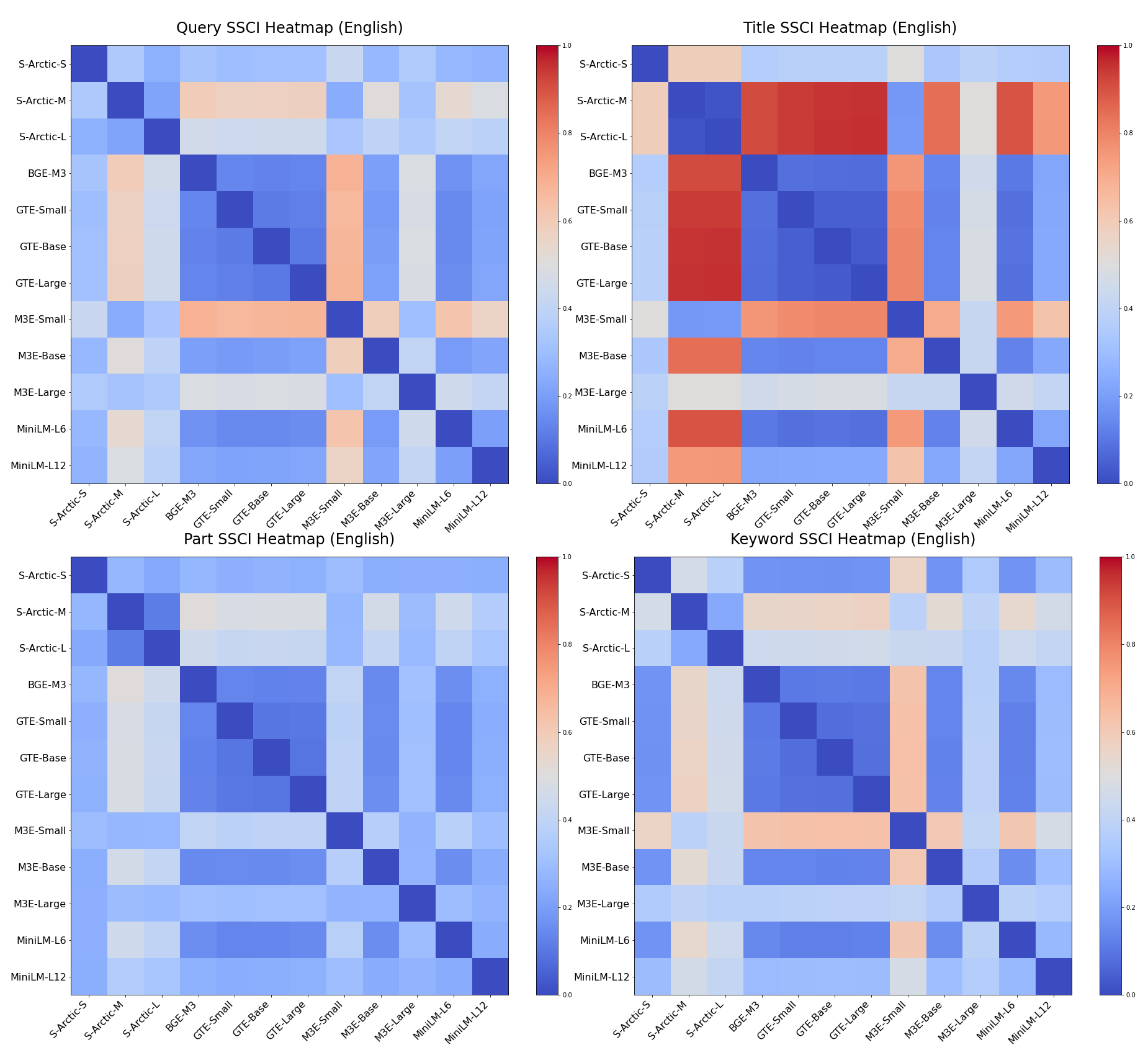}}
    \caption{Comparative SSCI Heatmaps of the S-Arctic Series, BGE Series, GTE Series, M3E Series, and MiniLM Series in the IRSC Benchmark's Query, Title, Part and Keyword Subtasks in English. Smaller values indicate more consistent model performance.}
    \label{figure:ssci2}
\end{figure}

Figure~\ref{figure:ssci2} illustrates the SSCI heatmaps for models on four tasks (excluding Summary) in the English language. From the figure \ref{figure:ssci2}, it is evident that different tasks exhibit varying degrees of divergence in model understanding. Unlike the Summary task, where most models show blue regions indicating high SSCI values, the IRSC tasks reveal different levels of red regions. This is particularly evident in the Title task, which shows extensive deep red regions, indicating significant divergence in semantic understanding among the models. The Title task imposes higher demands on the models' semantic understanding, highlighting the differences in model performance across different task types. While models show high consistency in the Summary task, possibly due to its clear objective and relatively smaller information processing requirements, the Title task requires complex context understanding and concise expression, leading to more pronounced divergences among the models. 

From the analysis of Figure~\ref{figure:ssci2}, we can derive that the Summary task shows high model consistency, whereas the IRSC tasks, especially the Title task, exhibit significant divergences, indicating that task complexity has a substantial impact on model consistency. Tasks requiring complex context understanding and concise expression (such as Title) reveal significant divergences in model performance, exposing limitations in these areas. Future model training should focus on enhancing models' capabilities in complex context understanding and concise expression to address the challenges posed by complex tasks.

\subsubsection{RCCI}

\begin{figure}[h]
    \centering
    \resizebox{0.95\linewidth}{!}{\includegraphics[]{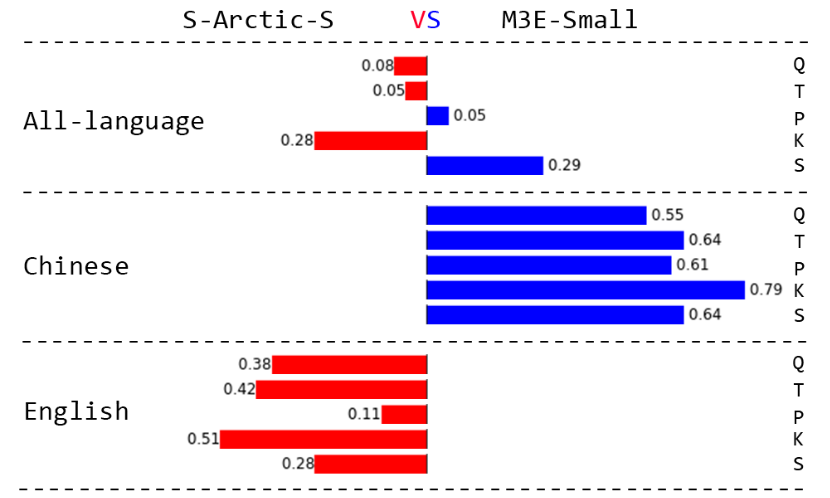}}
    \caption{Comparison of RCCI Results Between S-Arctic-S and M3E-Small Across Mixed-Languages, Chinese, and English}
    \label{figure:rcci}
\end{figure}

The figure \ref{figure:rcci} presents a comparative analysis of the capabilities of two models, S-Arctic-S and M3E-Small, across multiple languages and evaluation metrics. The RCCI methodology has been employed to offer a detailed comparison between the models, which overcomes the limitations of simple average-based metrics such as r@10, m@10, and n@10 by highlighting finer differences in model performance.

Figure~\ref{figure:rcci} illustrates the RCCI-based analysis clearly delineates the strengths and weaknesses of the two models. M3E-Small excels in the Chinese language, demonstrating robust performance across all metrics, which underscores its potential for applications requiring Chinese language proficiency. Conversely, S-Arctic-S exhibits a competitive edge in the English language metrics, particularly in the Q (Query), T (Title), and K (Keyword) metrics.

\section{Conclusion}

The IRSC Benchmark offers a comprehensive evaluation framework for embedding models in Retrieval-Augmented Generation (RAG) tasks. It includes five retrieval tasks: query-based, title-based, part-of-paragraph-based, keyword-based, and summary-based retrieval, in English and Chinese. Key contributions are the IRSC Benchmark, new metrics like the Similarity of Semantic Comprehension Index (SSCI) and Retrieval Capability Contest Index (RCCI), and a cross-lingual performance evaluation.

Experimental results show BGE-M3's superior performance across various metrics and tasks, highlighting its robust retrieval capabilities in monolingual and cross-lingual contexts. Diverse model performance emphasizes the importance of selecting models based on specific tasks and language needs. Cross-lingual retrieval challenges indicate a need for improved training in vector models for better semantic alignment. Visual tools like radar charts and heatmaps illustrate model strengths and weaknesses in different tasks and languages, showcasing the IRSC Benchmark's comprehensive evaluation.

Future research should focus on optimizing embedding models for complex tasks and improving cross-lingual semantic alignment.

\section{Limitations}

While the IRSC Benchmark provides a comprehensive evaluation framework for embedding models in Retrieval-Augmented Generation (RAG) tasks, there are several limitations that need to be addressed:

\begin{enumerate}
     \item \textbf{Language Scope}: The benchmark primarily focuses on English and Chinese, which limits its applicability to other languages. While it provides insights into multilingual capabilities, extending the evaluation to a broader range of languages would offer a more holistic view of model performance in truly multilingual settings.
     \item \textbf{Task Scope:}: Although the benchmark covers five distinct retrieval tasks, real-world RAG applications might involve more complex and diverse scenarios. Expanding the range of tasks to include more specialized or domain-specific queries could provide a more comprehensive assessment.
     \item \textbf{Model Variability}: The benchmark evaluates a selection of popular embedding models, but it does not encompass all existing models. New models and variations are continuously being developed, and the benchmark needs to be updated regularly to include these advancements.
     \item \textbf{Interoperability with Other Systems}: The benchmark does not assess how well these embedding models integrate with other systems and technologies used in RAG pipelines. Evaluating interoperability and integration efficiency could provide a more practical measure of model utility.

\end{enumerate}

Addressing these limitations in future research will help improve the robustness and applicability of the IRSC Benchmark, making it a more powerful tool for evaluating and developing embedding models in RAG tasks.

\bibliography{custom}

\clearpage

\end{document}